\documentclass[]{aa}
\input psfig.sty
\usepackage[english]{babel}
\usepackage[latin1]{inputenc}
\usepackage{graphicx}
\usepackage{psfrag}
\usepackage{subfigure}
\usepackage{rotating}
\usepackage{lscape}
\newcommand{\mb}[1]{\mbox{\boldmath$#1$}}     % kursiver Fettdruck in Formeln
\newcommand{\mr}[1]{\mathrm{#1}}              % für Indizes
\newcommand{\ind}[1]{\mr{#1}}                 % Indizes in Roman
\newcommand{\dif}[2]{\frac{\partial #1}%      % Partielle Ableitung
                          {\partial #2}}      %
\newcommand{\Dif}[2]{\frac{\mathrm{d} \/#1}%  % Totale Ableitung
                    {\mathrm{d} \/#2}}        % 
\renewcommand{\vec}[1]{\mb{#1}}               % ersetzt \vec durch \mb

\newcommand{\invh}[1]%                        % Setzt Hoehe des Arguments
   {\newlength{\hei}%
    \newlength{\dep}%
    \settoheight{\hei}{#1}%
    \settodepth{\dep}{#1}%
    \addtolength{\hei}{\dep}%
    \rule[\dep]{0pt}{\hei}%
   }
\newcommand{\invw}[1]%                        % Setzt Breite des Arguments
   {\newlength{\wid}%
    \settowidth{\wid}{#1}%
    \rule{\wid}{0pt}%
   }

\newcommand{\aref}[1]%
	{(\/\textit{Abb.~\ref{#1}}\/)} 	            % Abbildungsreferenz

\newcommand{\bref}[2]%
	{(\/\textit{Abb.~\ref{#1}(#2)}\/)}          % Abbildungsreferenz mit
\newcommand{\be}{\begin{equation}}
\newcommand{\ee}{\end{equation}}
\newcommand{\bea}{\begin{eqnarray}}
\newcommand{\eea}{\end{eqnarray}}

\begin{document}
\title{In situ acceleration in the Galactic Center Arc}
 \subtitle{}
 \author{S. Lieb, H. Lesch \and G.T. Birk}
 \offprints{S. Lieb}
 \institute{Institute for Astronomy and
   Astrophysics, University of Munich, Scheinerstr 1, D-81679 Munich,
   Germany}
 \date{}
 \mail{lieb@usm.uni- muenchen.de}

\abstract{For the nonthermal radio emission of the Galactic Center Arc
	in situ electron acceleration is imperative. The observed radio
	spectrum can be modeled by a transport equation for the relativistic
	electrons which includes particle acceleration by electric fields,
	momentum diffusion via scattering by magnetohydrodynamical
	turbulence and energy losses by synchrotron radiation. The
	accelerating electric fields can be regarded as a natural
	consequence of multiple reconnection events, caused by the
	interaction between a molecular cloud and the Arc region. The radio
	spectrum and even the recently detected 150 GHz emission,
	explicitely originating from the interaction regions of a molecular
	cloud with the magnetized Arc, can be explained in terms of
	quasi-monoenergetically distributed relativistic electrons with a
	typical energy of about 10 GeV accelerated in stochastically
	distributed magnetic reconnection zones.

}

\titlerunning{Acceleration in the Galactic Center Arc}

   \maketitle

%
%___________________________________

\section{Introduction}
The Galactic Center (GC) Arc is a unique nonthermal radio continuum
structure in the Galaxy. It is located at a projected distance of
about 30 pc from the Sgr A complex and consists of a network of
magnetic filaments filled with relativistic electrons running almost
perfectly perpendicular to the Galactic plane (e.g. \cite{Yus84}). The
nonthermal nature of these filaments has been proven beyond any doubt
by radio emission which exhibits a degree of linear polarization which
is close to the intrinsic value at high frequencies of 60\%
\cite{reich88, lesch92}. We emphasize the nonthermal character of
the Arc since its radio spectrum is a very remarkable one. A
decomposition of its radio spectrum between 843 MHz and 43 GHz by
\cite{reich88} shows a spectral index $\alpha = + 0.3$ (we use the
conventional relation that the observed flux $S_\nu$ scales as
$\nu^{\alpha}$). This finding of an increasing radio spectral index
has been confirmed by high resolution VLA-observations \cite{anan91}.

Such an inverted radiation spectrum with an index of +0.3 is expected
from a quasi-monoenergetic electron distribution or an energy
distribution with a well-defined low energy cutoff, respectively
\cite{lesch88}. Observations at high frequencies (between 32 GHz
\cite{lesch92} and 43 GHz \cite{sof87b}) seem to indicate a spectral
turn over, i.e. a fading of the Arc towards higher
frequencies. However, Reich et al. (2000) detected an enhanced
emission at 150 GHz slightly offset relative to the most intense
vertical nonthermal filaments seen at lower frequencies. This emission
originates from the apparent interacting areas of dense molecular
material with the Arc. 150 GHz emission was also detected south of a
molecular cloud, where the vertical filaments of the Arc cross a weak
filamentary structure.  The spectrum of the emission is inverted
relative to 43 GHz and compatible with an origin from
quasi-monoenergetic electrons or an electron distribution with a low
energy cutoff, but not with optically thin emission from cold
dust. Reich et al. (2000) conclude "that the coincidence of enhanced
emission with regions of interacting molecular gas strongly suggests
that high-energy electrons are accelerated in those places where the
magnetic field is compressed". They calculated a Lorentz factor of the
electrons to be $\gamma\simeq 2\times 10^4$ emitting synchrotron
radiation at 150 GHz within a magnetic field of 1mG.

To summarize, in the Galactic center several areas appear to be filled
with relativistic electrons whose energy must be distributed
quasi-monoenergetically around a few GeV.  We note that even the very
center of our Galaxy Sgr A$^{*}$ exhibits a radio to infrared spectrum
which is in surprisingly good agreement with optically thin
synchrotron emission of a quasi-monoenergetic electron distribution
with a typical energy of 80 MeV, i.e. Lorentz factor of
160 \cite{dusch94}.

In a former paper about particle acceleration in the Arc
\cite{lesch92} we considered the magnetohydrodynamical interaction of
a molecular cloud with the Arc filaments. Especially we investigated
the role of the moving gas cloud as a trigger mechanism for magnetic
field amplification accompanied by dissipation of the magnetic energy
via magnetic reconnection.

Magnetic reconnection takes place when magnetic field lines with
antiparallel directions encounter.  Such a situation corresponds to
the formation of an electric current sheet.  Typically astrophysical
plasmas are ideal electric conductors, i.e. their electrical
conductivity is very high. In such media the magnetic field is frozen
into the motion of the conducting fluid. Any plasma velocity
distortion onto the magnetic field lines is automatically related to
an electric field which is perpendicular to the plasma velocity and
the magnetic field. This electric field is described by the ideal
Ohm's law
\be
\vec{E} + \frac{1}{c} \vec{v} \times \vec{B} = 0.
\ee
In a plasma where the magnetic field is strongly distorted by shear
flows, radial explosive flows or stochastic motions, it is unavoidable
that field lines with antiparallel directions encounter. Consequently,
the ideal conducting plasma switches into a nonideal medium with a
finite, localized electrical conductivity, i.e. the ideal form of
Ohm's law is violated. The strong spatial gradients of the magnetic
field in such interaction zones represent an energetic "crisis" which
is relaxed by partial dissipation of the magnetic energy via the
formation of current sheets in the resistive medium (e.g. \cite{pri00}
and references therein). The nonideal Ohm's law is given by
\be
\vec{E} + \frac{1}{c}\vec{v} \times \vec{B} = \vec{R} \neq 0,
\ee
where $\vec{R}$ is some yet unspecified nonideal term, i.e. the plasma
resistance.

In other words, magnetic reconnection corresponds to a magnetic field
aligned electric field which is associated with a generalized electric
potential \cite{sch88, sch91}
\be
V=-\int{E_s \; \rm{ds}}=-\int{R_s \; \rm{ds}}
\ee
where the integral is evaluated along the magnetic field lines that
penetrate the reconnection region. Of course, such a potential drop offers the
possibility to accelerate particles very efficiently \cite{sch91,
  black96, lesch97, lesch98, sch98, lit99, nod03}.

Instead of investigating the acceleration in one reconnection zone, we
consider in this contribution the acceleration of relativistic
electrons in numerous reconnection regions, i.e. current carrying
filaments driven by the interaction of a molecular cloud with the
magnetic field in the Galactic Center Arc \cite{lesch92, ser94}.  This
scenario arises quite naturally, since the necessary energy source is
represented by moving molecular gas which encounters the poloidal
magnetic fields in the Arc which is proven to be present by the
observed very high polarization of the radio emission up to 60\%.

Since we have many acceleration regions we apply a dynamical
description of the energy distribution function of relativistic
electrons based on systematic momentum gains by reconnection, losses
by synchrotron losses and momentum diffusion due to scattering on
magnetohydrodynamical turbulence, i.e. Alfv\'en waves and magnetosonic
waves.

\section{The temporal evolution of the energy distribution function of relativistic particles}

The dynamical evolution of a population of high-energy particles that
do not interact with each other is described by the Liouville equation

\be
\dif{F}{t} + \Dif{\bf r}{t} \nabla_{\bf r} F + \Dif{\bf p}{t}
\nabla_{\bf p} F = 0
\label{1}
\ee
where $F({\bf r}, {\bf p}, t)$ is the one-body distribution function.
Due to the complexity of the non-linear dynamics in 6+1 dimensions
this equation is only of limited use.  Fortunately, for our purposes
we do not need to know the full information of the particle dynamics.
Rather, we will make use of the following assumptions.  First, we will
assume pitch angle isotropy.  This should be granted by efficient
particle scattering in momentum space in reconnection regions caused
by Alfv\'enic and magnetosonic wave fields with relatively high-energy
densities (see \cite{sch86}).  Second, we are interested in the global
energy spectrum of the radiating electrons and thus, can dispense with
the detailed spatial dependence of the distribution function.

Consequently, we deal with the isotropic distribution $f(p,t)=\int d
{\bf r} F({\bf r}, p, t)/V $ where $V$ denotes the emission volume.  
The total particle number is calculated as $N(t)=4\pi\int f
p^2dp$.  The continuous momentum gains and losses $\Delta p$ of the
particles can be described by the momentum operator

\be
{\cal L}_p = \frac{1}{p^2}\dif{}{p}[(\dot p_\ind{gain}+ \dot
p_\ind{loss})p^2]\label{5},
\ee
where the systematic temporal momentum gains and losses are denoted by
$\dot p_\ind{gain}$ and $\dot p_\ind{loss}$, respectively.

Additionally, we will allow for stochastic diffusion in momentum space by 
scattering of magnetohydrodynamical fluctuations
(in the sense of a Fokker-Planck term) denoted by $D_p$.  In
principle, catastrophic losses that may result to a further sink in
the balance equation for $f$ can be modeled by a characteristic time
$\tau$, i.e.  $\partial f/\partial t \sim - f/\tau$.  Finally, the
injected particle population is represented by some explicit source
term $S(p)$.  Accordingly, the temporal evolution of the particle
distribution is governed by the equation \cite{sch84, sch86}

\be
\dif{f(p, t)}{t} - \frac{1}{p^2}\dif{}{p}\left(D_p p^2 \dif{f(p,
t)}{p}\right) + {\cal L}_p f(p, t) + \frac{f(p, t)}{\tau} =
S(p)\label{4}
\label{eq:fte}
\ee
which holds, if the distribution function $f({\bf r}, p, t)$ is
separable in space and momentum, and if the momentum operator ${\cal
L}_p$ does not depend on the spatial coordinates.  In the present
application based on the reconnection scenario (see
also \cite{bir01}) the momentum gain is
given by $\dot p_{gain}=\frac{d}{dt}(\gamma m v) = q E_\ind{s} =
\zeta={\rm const.}$, where $\gamma$ denotes the Lorentz factor,
$E_\ind{s}$ is the accelerating electric field component (locally directed
parallel to the magnetic field) and $q$ is the electrical charge.  

We emphasize that the acceleration of high-energy particles is
regarded as a consequence of the conversion of magnetic field energy
into particle energy by magnetic reconnection.  Since synchrotron
losses are responsible for the observed non-thermal radio emission
from the Arc we take $\dot p_{loss}=-kp^2$ with
$k=\frac{4}{3}\sigma_\ind{T}U_\ind{B}/m_\ind{e}^2c^2$, with the
Thomson cross section $\sigma_\ind{T}$ , the electron mass
$m_\ind{e}$, the speed of light $c$ and the magnetic field energy
density $U_\ind{B}$.  For our applications inverse Compton
scattering, which has the same $\gamma$-dependence as synchrotron
losses, as well as bremsstrahlung are negligible.  In case of particle
scattering in momentum space by Alfv\'enic waves the momentum
diffusion coefficient is given \cite{sch84} by $D_p = Dp^2$ with $D =
v_\ind{A}^2/9K(p)$, where $v_\ind{A}$ is the Alfv\'en velocity and
$K(p) = \lambda c\beta/3$ is the spatial diffusion coefficient, which
has been expressed in terms of the particle's mean free path $\lambda$
\cite{sch86}. For our purpose we have set $\beta = v/c = 1$. Since we
have to model the formation of monoenergetic distribution functions,
the effect of catastrophic losses must be negligible, because
otherwise the particles would leave the acceleration region before a
pile up in energy has been established \cite{sch86}.

\section{Numerical results}
In order to calculate the temporal evolution of $f(p, t)$ by numerical
integration of equation 6 we first have to specify the physical
parameters $\zeta$, $k$, $D$, and $S(p)$, which are related to
the physical conditions of the Galactic Center Arc.  First of all, we
discuss the source term $S(p)$.  Since the energy distribution of the
injected particles, which are generated in the Galactic center, is a
monoenergetic one \cite{lesch92}, the use of a $\delta$-like source
term, represented, e.g., by a $cosh^{-4}(p)$-function in our
calculations, is a reasonable assumption. As a typical injection
energy we use 75 MeV, the value close to the one calculated by
\cite{dusch94} for the central object Sgr A${^*}$, i.e. $\gamma_{inj}
= 150$.

The magnetic field strength in the arcs filaments is of the order of a
few mG, or higher \cite{Yus87}.  To be on the save side we use the 150
GHz detection of \cite{reich00} close to the Arc as the maximum
synchrotron frequency emitted.  Thus, we have to deal with a Lorentz
factor of about $\gamma \simeq 2\cdot10^4$, necessary for 150 GHz
synchrotron emission in a magnetic field of $10^{-3}$ G.

The maximum electric field strength is given by $E\simeq w/c B$, where
$w$ denotes the bulk velocity of the thermal plasma that acts as an
MHD generator.  The
detected cloud motion is in the range of 15-45 km/s
\cite{tsu97}.  Again to be on the save side we will use $v= 5$ km/s
which gives a maximum electric field strength of $10^{-8} statvolt \,
cm^{-1}$.  If we assume, that electrons can be accelerated over the
whole extension of the Arc (about 40 pc) without energy losses (ideal
linear accelerator), we need a minimum strength of the average
electric field of $E_\ind{s} \simeq 3 \cdot 10^{-13} \; \rm{statvolt
\, cm^{-1}}$ to reach the observed particle energies, which is a tiny
fraction of the available electric field.

This oversimplification is helpful to work out the analysis with a
reference value of $\zeta$ and is not meant as a physical assumption
we apply (see discussion below).

If we restrict our calculations to electrons we obtain an estimate for
the systematic acceleration term
\be
\dot p_\ind{gain} = eE_\ind{s} = \zeta \ge 1.5\cdot 10^{-22} \;
\rm{dyne}.
\ee
Due to the high polarization of the synchrotron radiation, up to 60 \%
\cite{lesch92} we consider only synchrotron emission to be responsible for
the energy losses of the particles in the Arc.  In this case we can find
a certain value $p_\ind{0} = ( 3\zeta m_\ind{e}^2 c^2/4\sigma_\ind{T}
U_\ind{B} )^{1/ 2}$ where momentum gains and losses keep the balance.
For momenta $p > p_\ind{0}$ momentum loss dominates momentum gain.  For
momenta $p < p_\ind{0}$ the opposite holds.  Thus, we expect a pile
up for the distribution function $f(p, t)$ at $p_\ind{0}$, which is
consistent with a monoenergetic distribution of the electrons (see
also \cite{lesch92}).  As discussed above, the pile up in the Arc
should occur at momenta with Lorentz factor $\gamma\simeq 2\cdot
10^4$.  If one assumes, for simplicity, a constant electric field of
$E_\ind{s} \simeq 3 \cdot 10^{-13} \; \rm{statvolt \, cm^{-1}}$ over
the complete expansion of the arc, a magnetic field of $3\cdot10^{-3}$
Gauss is necessary for a pile up at $p_\ind{0} = \gamma m_\ind{e}c =
5.4\cdot 10^{-13} \; \rm{g\,cm\,s^{-1}}$ according to a systematic
loss coefficient of
\be
k = \frac{4}{3}\sigma_\ind{T}\frac{U_\ind{B}}{m_ \ind{e}^2c^2} = 1.4\cdot
10^3\;\rm{g^{-1}\,cm^{-1}}.
\ee
However, the particles, in fact, will be accelerated locally in
numerous reconnection regions with $E_{\vert \vert}\ne 0$ and will be
scattered by magnetohydrodynamic fluctuations like Alfv\'enic and/or
magnetosonic waves.  Consequently, the electric field strength in the
reconnection regions must be ultimately stronger than $10^{-13} \;
\rm{statvolt \, cm^{-1}}$.  Indeed, the electric field can be
estimated by the ratio of the number density of relativistic electrons
$n_\ind{r}$ and the thermal electron density $n_\ind{e}$ \cite{papa77}
\be
\frac{n_\ind{r}}{n_\ind{e}} \simeq \exp{\left[ -\frac{E_\ind{c}}{2E}
    \right]}.
\ee
$E_\ind{c} \simeq m_\ind{e}v_\ind{the}\nu_\ind{coll}/e$ denotes the
critical electric field, whereas $v_\ind{the}$ is the thermal velocity
of the electrons and $\nu_\ind{coll}$ is the collision frequency.  As
calculated by Lesch \& Reich 1992 the thermal electron number density
$n_\ind{e}$ in the Arc filaments is about $10\;\rm{cm}^{-3}$ and the
density of relativistic electrons is of the order of
$10^{-7}\;\rm{cm}^{-3}$. If we use an electron temperature of $10^4$ K
 we obtain from equation 9  $E
\simeq 3\cdot10^{-10} \; \rm{statvolt \,cm^{-1}}$, which can be easily
achieved by the available electric field, induced by the
$\vec{w}\times\vec{B}$ - motion of a molecular clouds calculated
above.

To obtain the diffusion coefficient $D$ we first have to estimate
the particle's mean free path $\lambda$.  As shown in \cite{sch02}
$\lambda$ can be determined by
\be
\lambda \simeq r_\ind{G}\left( \frac{B_\ind{0}}{\delta B} \right)^2
\left( \frac{L_\ind{0}}{r_\ind{G}} \right)^{q-1}
\ee
where $r_\ind{G}$ is the gyro radius of the protons, $B_\ind{0}/\delta
B$ gives the ratio between macroscopic and fluctuated magnetic field
strength, $L_\ind{0}$ is the expansion of the considered object and
$q$ is turbulence spectral index.  In the vicinity of a strong
magnetic distortion, like the Arc filaments, Burgers turbulence is
expected to be excited \cite{cham88}. Thus, we choose a turbulence
spectral index of $q=2$. With a
particle density of $n \sim 10\;\rm{cm}^{-3}$ and a magnetic field
strength of $3\cdot10^{-3}$ Gauss the Alfv\'enic velocity is about
$700\;\rm{km\;s}^{-1}$.  Hence we obtain for the diffusion coefficient
\be
D = \frac{v_\ind{A}^2}{3c\lambda} \simeq \frac{1}{3}\left(
\frac{\delta B}{B_\ind{0}} \right)^2 \frac{v_\ind{A}^2}{cL_\ind{0}} = 5.4\cdot
10^{-18}\rm{s^{-1}},
\ee
if we assume, that $B_\ind{0}/\delta B$ is of the order of 10 which
implies a reasonable level of well-developed turbulence. Much higher
diffusion coefficients would inhibit pile-up distributions. However, a
quasi-linear theory applied for equation 9 requires $\delta B < B_\ind{0}$.

The synchrotron emission spectrum associated with the particle
spectrum can be calculated from \cite{ryb79}

\be
I_{\nu}=\int P_\ind{\nu}(p)N(p)\,\mr{dp},
\ee
where

\be
P_{\nu}(p)=\frac{2\sqrt{3}}{3}\frac{ e^2m_\ind{e}^{2}c\nu}{p^2}
\int_{\nu/\nu_\ind{c}}^\infty K_\ind{\frac{5}{3}}(\xi)\,\mr{d\xi},
\ee
is the power per unit frequency $\nu$ emitted by each electron.
$K_\ind{\frac{5}{3}}(\xi)$ denotes the modified Bessel function of the
second kind and $\nu_\ind{c} = 3\gamma^3 \omega_\ind{G}/4\pi $ is the
critical frequency ($\omega_\ind{G}$ is the electron gyro frequency).

Having clarified the initial conditions and chosen physical parameters
we now come to the numerical results.  Figure 1 shows the temporal
evolution of the particle density $N(\gamma)$ and the corresponding
synchrotron emission spectrum, calculated from equation 12. In the
chosen normalization the time is measured by $t_0=10^9$ s.  As long as
the particle acceleration exceeds the synchrotron losses, the
quasi-monoenergetic distribution of the electrons is shifted toward
higher momenta (a, c). Due to momentum diffusion the maximum of
$N(\gamma)$ decreases with time (c). As soon as the synchrotron losses
are comparable to the energy gains, the maximum of $N(\gamma)$
increases again (Figure 2 (a)).  At a Lorentz factor of about $\gamma
= 2\cdot10^4$ the momentum gains and losses are balanced and the
distribution of the electrons that are assumed to be contained in the
Arc during the entire radiation processes develops towards a shifted
pile up. Consequently, a monoenergetic distribution function evolves
again (Figure 2 (c)).  The cut-off frequency remains almost constant
during entire pile-up dynamics.  The associated radiation spectrum
(Figure 2 (c)) corresponds to the one observed by several authors,
including the 150 GHz emission detected by \cite{reich00}.  We
note that for a finite loss time of $\tau \le 10^{10}$ s the
pile-up is prevented by the particle losses.

\section{Conclusions}
In the filaments of the Galactic Center Arc electrons are accelerated
to considerably high energies of about 10 GeV. This is obvious from
the detected polarized radio emission exhibiting a very high degree of
polarization (60\%) and a rising spectrum with a spectral index of
+0.3 up to 150 GHz. Such a spectral behavior is due to an energy
distribution function of the radiating relativistic electrons which
either has a low-energy cutoff or which is quasi mononenergetic. No
obvious energy sources for particle acceleration are present in the
neighborhood of the filaments. The very center of the galaxy,
SgrA${^*}$ is a source for monoenergetic relativistic particles but on
significant lower energies of about 50 MeV \cite{dusch94}. However,
the magnetic filaments interact with the plasma of a molecular cloud
which moves with velocities of some 10 km ${\rm s}^{-1}$ relative to
the magnetic field.  This interaction can be interpreted in terms of
the induced Lorentz force inducing a convective electric field which
in the first place is oriented perpendicular to the magnetic field and
the local plasma velocity of a molecular cloud.  The motion of the
plasma locally distorts the magnetic field necessarily in such a way
that antiparallel directed field lines encounter. Provided that some
violation of ideal Ohm's law occurs, e.g. due to microturbulence,
magnetic reconnection will occur. The magnetic field energy will be
partly converted to particle energization. In the reconnection regions
a magnetic field-aligned electric field component forms with a
magnitude of some fraction of the convective electric field.  This
parallel electric field represents a perfect candidate for efficient
particle acceleration. The required electric field for the electron
energization are considerably weaker than the magnetic field which is
in accordance with the observational fact that the cloud plasma moves
with velocities much lower than the speed of light.  Since
reconnection is a localized phenomenon which depends on the local
properties of the plasma, we consider a scenario characterized by
multiple reconnection sheets in which particles are accelerated and
scattered by magnetohydrodynamical fluctuations.  Such scattering
leads to diffusion in energy space.  Moreover, we take into account
the energy losses by the observed synchrotron radiation.  By means of
a relatively simply transport equation for the distribution function
of the radiating relativistic electrons we could show that an injected
monoenergetic distribution function with 75 MeV (coming from the
Galactic center) evolves into a quasi-monoenergetic distribution
function at 10 GeV.  The resulting radiation spectra increase with
frequency up to some hundred GHz according to the observations,
i.e. $S_\ind{\nu} \sim \nu^{1/3}$.

A major assumption of our study is the complete isotropy of the
distribution function which is based on calculations by \cite{acha90},
who show that the isotropisation time $t_\ind{iso} =
c/v_\ind{A}(B/\delta B)1/\Omega$, where $\Omega$ is
the gyro frequency of the protons. In this application $t_\ind{iso}$ is
about $4\cdot10^6$ s which is much shorter than the studied time scale
of the evolution of the distribution function. Some non-isotropic
particle population would lead to an overlap of a second radiation
component in a specific energy range that we can not handle in our
model. However, the present observations seem to give no hint to such
a component.

\begin{figure}
\subfigure[Temporal evolution of $N(\gamma)$. The solid, dotted and
dashed lines show the injected particle spectrum and the evolution
after $t=1\,t_0$ and $t = 2\,t_0$, respectively.]{
\includegraphics[angle=0,width=0.55\linewidth,keepaspectratio]{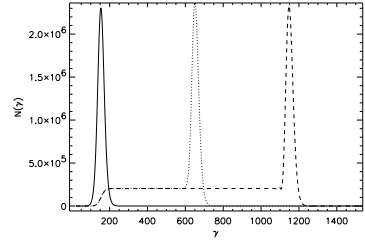}
}
\subfigure[The corresponding synchrotron spectrum $I_{\nu}$ ]{
\includegraphics[angle=0,width=0.55\linewidth,keepaspectratio]{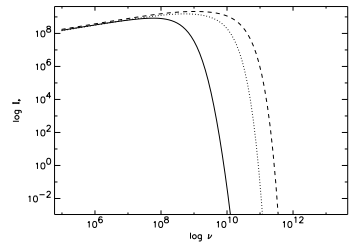}
}
\subfigure[Temporal evolution of $N(\gamma)$. The solid and
dotted lines show the evolution after $t=10\,t_0$ and $t=20\,t_0$,
respectively.]{
\includegraphics[angle=0,width=0.55\linewidth,keepaspectratio]{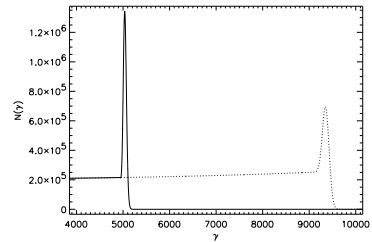}
}
\subfigure[The corresponding synchrotron spectrum $I_{\nu}$]{
\includegraphics[angle=0,width=0.55\linewidth,keepaspectratio]{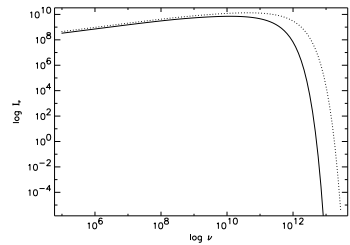}
}
\caption[] {Temporal evolution of $N(\gamma)$ calculated from equation
	6. The values of the physical parameters, which we have used for
	our calculations are: $\zeta = 1.5\cdot 10^{-22}$ dyne, $k =
	1.4\cdot 10^3\;\rm{g^{-1}\,cm^{-1}}$ and $D = 5.4\cdot 10^{-18}
	\rm{s^{-1}}$. To evaluate the synchrotron emission spectrum 
                    (see equation 12) we have
	used a magnetic field strength of $B = 3$ mG.}
\label{}
\end{figure}

\begin{figure}
\subfigure[Temporal evolution of $N(\gamma)$. The solid and
dotted lines show the evolution after $t=30\,t_0$ and $t=50\,t_0$,
respectively.]{
\includegraphics[angle=0,width=0.55\linewidth,keepaspectratio]{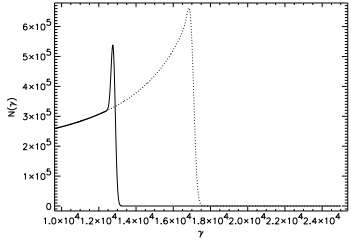}
}
\subfigure[The corresponding synchrotron spectrum $I_{\nu}$]{
\includegraphics[angle=0,width=0.55\linewidth,keepaspectratio]{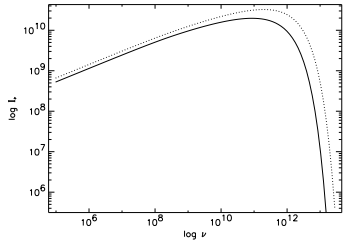}
}
\subfigure[Temporal evolution of $N(\gamma)$. The solid, dotted and
dashed lines show the evolution after $t=80\,t_0$, $t=100\,t_0$ and
$t=150\,t_0$, respectively.]{
\includegraphics[angle=0,width=0.55\linewidth,keepaspectratio]{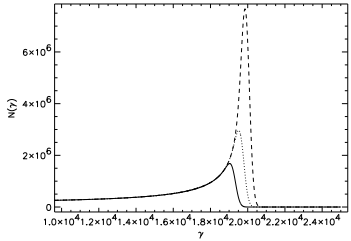}
}
\subfigure[The corresponding synchrotron spectrum $I_{\nu}$]{
\includegraphics[angle=0,width=0.55\linewidth,keepaspectratio]{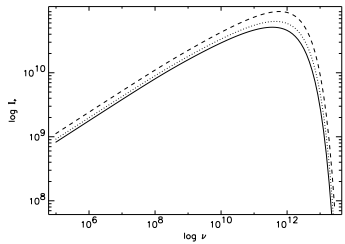}
}
\caption[]
{Temporal evolution of $N(\gamma)$ towards a quasi-monoenergetic
distribution function (a), (c). The resulting synchrotron emission
spectrum (d) shows an inverted spectrum with a spectral index of $\alpha
= 1/3$ ($I_\nu \sim \nu^{\alpha}$) and a cutoff-frequency of some
hundred GHz after $1.5\cdot10^{11}$ s.}
\label{fig2}
\end{figure}

\end{document}